\documentclass[useAMS,usenatbib,usegraphicx]{mn2e}
\onecolumn{}

\title[NY Virginis and its Circumbinary Planets]{The Pulsating sdB+M Eclipsing System NY Virginis and its Circumbinary Planets}

\author[Jae Woo Lee et al.]
       {Jae Woo Lee$^{1,2}$\thanks{E-mail: jwlee@kasi.re.kr (JWL); tchinse@gmail.com (TCH); jhyoon@kasi.re.kr (J-HY); whan@kasi.re.kr (WH)},
        Tobias Cornelius Hinse$^{1,3}$\footnotemark[1], Jae-Hyuck Youn$^1$\footnotemark[1] and Wonyong Han$^{1,2}$\footnotemark[1] \\
        $^1$Korea Astronomy and Space Science Institute, 776 Daedeokdae-ro, Yuseong-gu, Daejon 305-348, Korea\\
        $^2$Astronomy and Space Science Major, Korea University of Science and Technology, Daejeon 305-350, Korea \\
        $^3$Armagh Observatory, College Hill, BT61 9DG, Armagh, UK }

\begin{document}

\date{Accepted 2014 ---------. Received 2014 ---------; in original form 2014 }

\pagerange{\pageref{firstpage}--\pageref{lastpage}} \pubyear{2011}

\maketitle

\label{firstpage}

\begin{abstract}
We searched for circumbinary planets orbiting NY Vir in historical eclipse times including our long-term CCD data. 
Sixty-eight times of minimum light with accuracies better than 10 s were used for the ephemeris computations. The best fit 
to those timings indicated that the orbital period of NY Vir has varied due to a combination of two sinusoids with periods of 
$P_3$=8.2 yr and $P_4$=27.0 yr and semi-amplitudes of $K_3$=6.9 s and $K_4$=27.3 s, respectively. The periodic variations most
likely arise from a pair of light-time effects due to the presence of third and fourth bodies that are gravitationally 
bound to the eclipsing pair. We have derived the orbital parameters and the minimum masses, $M_3 \sin i_3$ = 2.8 M$\rm_{Jup}$ 
and $M_4 \sin i_4$ = 4.5 M$\rm_{Jup}$, of both objects. A dynamical analysis suggests that the outer companion is less likely 
to orbit the binary on a circular orbit. Instead we show that future timing data might push its eccentricity to moderate values 
for which the system exhibits long-term stability. The results demonstrate that NY Vir is probably a star-planet system, 
which consists of a very close binary star and two giant planets. The period ratio $P_{3}$/$P_{4}$ suggests that 
a long-term gravitational interaction between them would result in capture into a nearly 3:10 mean motion resonance. 
When the presence of the circumbinary planets is verified and understood more comprehensively, the formation and evolution 
of this planetary system should be advanced greatly.
\end{abstract}


\section{Introduction}

The presence of a third body causes the relative distance of the eclipsing pair to the observer to change as it orbits 
the barycenter of the triple system. The light-travel-time (LTT) effect produces a periodic variation in times of conjunction,
which can be studied by searching for quasi-sinusoidal features in $O$--$C$ diagrams (observed minus calculated minimum epochs).
Eclipse times have been exploited since the late 18th century and their accuracies have increased significantly with the use 
of modern observations with high quantum efficiency. The timing method offers us an opportunity to detect circumbinary 
substellar companions, such as giant planets and brown dwarfs, for eclipsing binaries with timing accuracies of a few seconds, 
if their historical database is large and sufficiently long (Lee et al. 2009a; Pribulla et al. 2012). Also, it should be 
possible to find additional planets or exomoons by analyzing transit timing variations in systems with known transiting planets 
(Doyle \& Deeg 2004; Holman 2005). The discovery and interpretation of such objects is of great interest because they can help 
in understanding the diverse architectures of planetary systems.

In order to search for low-mass objects orbiting a close binary using timing measurements, we have been observing 
eclipsing stars, primarily HW Vir-type detached systems which consist of a subdwarf B (sdB) primary and a normal main-sequence 
M companion with orbital periods between 2 and 4 hours. In our first result, we discovered two substellar companions revolving 
around the prototype HW Vir with orbital periods of 15.8 yr and 9.1 yr and having minimum masses of 19.2 $M\rm_{Jup}$ and 
8.5 $M\rm_{Jup}$, respectively, based on our photometric observations from 2000 through 2008 (Lee et al. 2009a). 
The circumbinary objects are the first planetary system unequivocally discovered using eclipse timings. 

NY Vir (PG 1336-018, GSC 4966-491, 2MASS J13384814-0201491; $V$=13.3) has been included as an object in our survey. 
The variable was recognized to be an HW Vir-like eclipsing binary of about 2.4 h period with a rapid sdB pulsator by 
Kilkenny et al. (1998). They determined the effective temperatures of $T_1$=33,000 K and $T_2$=3,000 K, the gravity of 
$\log g_1$=5.7, an orbital inclination of $i$=81$^\circ$, and the relative radii of $r_1$=0.19 and $r_2$=0.205. 
Since then, the system has been the target of a Whole Earth Telescope campaign (Kilkenny et al. 2003) and the subject of 
several investigations (Kilkenny et al. 2000; Hu et al. 2007; Vu\v ckovi\'c et al. 2007, 2009), which have been reviewed 
in a recent paper by Kilkenny (2011). It is known that the primary component is a rapidly pulsating sdB star exhibiting 
short-period multimode light variations and the secondary is a fully convective M dwarf. In this article, we present 
long-term CCD photometry of NY Vir and describe a search for circumbinary planets in this system from all available data. 
We then test the dynamical stability for our two-planet model and find several regions resulting in a stable orbital 
configuration over $10^6$ yr.

\section{New Long-term Observations and Eclipse Times}

New CCD photometric observations of NY Vir were taken on 17 nights from 2011 January through 2014 May in order to collect 
additional eclipse timings. We used CCD cameras and $I$ filters attached to the 1.0-m reflector at Mt. Lemmon Optical Astronomy 
Observatory (LOAO) in Arizona, USA. A summary of the observations is listed in Table 1, where we present observing interval, 
numbers of nights, CCD type, field of view (FOV), and numbers of observed points. The instruments and reduction methods for 
the FLI IMG4301E and ARC 4K CCD cameras are the same as those described by Lee et al. (2009b, 2012b). In order to 
make an optimal artificial comparison source during our observing runs, we monitored a few tens of stars imaged on the chip 
at the same time as the eclipsing pair. Following the procedure described by Lee et al. (2011), five useful field stars were 
selected and combined by a weighted average. The difference magnitudes between the variable and the artificial reference star 
were computed and the resultant light curves are displayed in Figure 1.

Over all observing seasons, 2,370 individual observations were obtained and a sample of them is listed in Table 2, where 
times are Barycentric Julian Dates (BJD) in the Barycentric Dynamical Time system (Eastman et al. 2010). From the LOAO 
observations, times of minimum light and their errors were determined using the method of Kwee \& van Woerden (1956). 
Forty-three new timings are given in Table 3; four of them are the weighted means calculated by us from 
the individual measurements of Vu\v ckovi\'c et al. (2007) in the $r'g'$ bandpasses of the SDSS system.

\section{Eclipse Timing Variation and its Implications}

Period studies of NY Vir have already been presented several times and the possible causes of the period variability have 
been discussed. From a quadratic fit, Kilkenny (2011) and \c Camurdan, \c Camurdan \& \. Ibano\v{g}lu (2012) showed that 
the period is decreasing at a rate of $-$4.1$\times$10$^{-9}$ d yr$^{-1}$. The latter authors reported that the period 
decrease may be explained by an angular momentum loss (AML) from this system or may be only a part of a cyclic variation. 
Most recently, from the analysis of 48 eclipse timings, Qian et al. (2012) found that the orbital period has varied 
through a combination of a downward parabola ($dP$/$dt$=$-$3.36$\times$10$^{-9}$ d yr$^{-1}$) and a sinusoid, 
with a period of 7.9 yr and a semi-amplitude of 6.1 s. They suggested that the sinusoidal variation could be produced 
by an LTT effect due to a circumbinary planet in this system with a minimum mass of 2.3 $M\rm_{Jup}$ and the parabolic change 
may be a part of an additional period modulation due to the possible presence of a fourth object.

For ephemeris computations of NY Vir, our measurements were added to the compilation used by Kilkenny (2011) and 
the eclipse timings observed by \c Camurdan, \c Camurdan \& \. Ibano\v{g}lu (2012) and Qian et al. (2012). First of all, 
we represented the minimum epochs by a parabolic least-squares fit. However, the quadratic ephemeris is not sufficiently 
detailed to fit all observed timings and the $O$--$C$ residuals from it seem to oscillate with a very small semi-amplitude 
($<$ 0.0001 d). This implies that the period change of the eclipsing pair could be explained by a combination of at least 
two causes, a long-term period decrease and a short-term oscillation. In order to examine a circumbinary companion in 
the system as the possible cause of the periodic variation, it is better to use LTT rather than the pure sine curve used 
by Qian et al. (2012). Such an ephemeris may be represented as:
\begin{eqnarray}
C_{1} = T_0 + PE + AE^2 + \tau_{3},
\end{eqnarray}
where $\tau_{3}$ symbolizes the LTT due to a third body (Irwin 1952). Because the secondary eclipses are much shallower 
and their timings less determinate than the primary ones, no secondary eclipses were used in our timing analyses. 
In addition, we excluded primary eclipse times with quoted errors larger than 0.0001 d. For the timings without errors 
published by Kilkenny (2000), their standard deviation of $\pm$0.00003 d was calculated to provide a mean error. Weights 
for all timings were then scaled as the inverse squares of their errors.

The Levenberg-Marquart (LM) algorithm (Press et al. 1992) was applied to evaluate the eight parameters of the ephemeris 
(Irwin 1959). Initial parameter values were obtained from Qian et al. (2012) and a periodogram analysis for 
the timing data and orbital eccentricities between 0 and 1 were examined. The results are summarized in 
the second column of Table 4, together with physical parameters derived by applying the absolute parameters 
($M_1$=0.466 M$_\odot$, $M_2$=0.122 M$_\odot$, $R_1$=0.15 R$_\odot$, $R_2$=0.16 R$_\odot$, $L_1$=19.3 L$_\odot$, and 
$L_2$=0.003 L$_\odot$) from Model II in Table 3 of Vu\v ckovi\'c et al. (2007). The LM technique produces 
the formal errors computed from the best-fit covariance matrix. Thus, we determined the parameter errors from 
Monte Carlo bootstrap-resampling experiments (Press et al. 1992, Hinse et al. 2014a). For this, we generated 
50,000 bootstrap-resampling datasets and used the best-fit model as our initial guess for each dataset. 
The parenthesized numbers in Table 4 are the 1$\sigma$-values associated with the last digits of the evaluations. 
We use the best-fit values as the average to compute the standard deviation from the 50,000 element bootstrap array.
The orbital period of the third body in an eccentric orbit of $e_3$=0.46 is somewhat longer than that determined by 
Qian et al. (2012). In the top panel of Figure 2, the solid and parabolic curves represent the full contribution and 
the quadratic term of the equation, respectively. The middle panel refers to the LTT orbit with a cycle length of 8.42 yr 
and the bottom panel displays the residuals from the full ephemeris. The quadratic {\it plus} LTT ephemeris gives 
a satisfactory fit to the mean trend of the timing residuals and results in a smaller $\chi^2_{\rm red}$=1.31 than 
the parabolic ephemeris ($\chi^2_{\rm red}$=2.08), although even $C_1$ does not fit well with the 2014 timings.

The downward parabolic curve in Figure 2 cannot be explained by a mass transfer between the binary components, because 
the eclipsing pair is a very detached system and both stars are close to spheres. Hence, the long-term period decrease at 
a rate of $-$3.36$\times$10$^{-9}$ d yr$^{-1}$ can be interpreted as AML corresponding to $dJ$/$dt$ = $-$1.38$\times$10$^{35}$ 
cgs units. Possible mechanisms for AML are gravitational radiation (Paczynski 1967) and/or magnetic braking in 
the cool secondary (Rappaport, Verbunt \& Joss 1983). We computed the AML rate for each mechanism in a procedure identical 
to that for HW Vir (Lee et al. 2009a). The theoretical rate for gravitational radiation is $-$1.01$\times$10$^{33}$, 
which is about two orders of magnitude smaller than the observed value, while the AML rate for the latter is 
$-2.07\times$10$^{35}$ for $\gamma =$ 2, all in cgs units. Added to our previous evaluation for HW Vir, this result makes 
it likely that the secular period decreases of sdB+M eclipsing binaries are mainly produced by AML due to magnetic stellar 
wind braking. However, because the secondary component should be a fully convective star, it may be that the magnetic braking 
may not be a dominant contributor to the observed period decrease of NY Vir. 

Another mechanism may be needed to explain the secular change. It is alternatively possible that the quadratic term can be 
just the observed part of a second LTT effect due to the existence of a more distant circumbinary object. Accordingly, 
the times of minimum light were fitted to a two-LTT ephemeris, as follows:
\begin{eqnarray}
C_{2} = T_0 + PE + \tau_{3} + \tau_{4}.
\end{eqnarray}
The LM method was used again in order to simultaneously evaluate the LTT parameters of the third and newly assumed fourth 
bodies. In this process, we have explored a large number of initial guesses for the LM method. Final results are 
given in the third and fourth columns of Table 4. We found that the parameter errors determined from the LM algorithm are 
too optimistic and underestimated and, thus, determined the parameter errors from 50,000 bootstrap-resampling simulations 
in a manner similar to that for the quadratic {\it plus} LTT ephemeris. The short-term periods for the two ephemerides 
agree with each other within the limits of their errors. The $O$--$C_2$ diagram with respect to the two-LTT ephemeris is 
drawn at the uppermost panel of Figure 3 with the solid curve due to the full effect of equation (2). The second and 
third panels display the $\tau_3$ and $\tau_4$ orbits, respectively, and the bottom panel shows the residuals from 
the complete ephemeris. The observed scatters seen in the $O$--$C_{\rm 2,full}$ residuals may be affected by 
the rapid multimode pulsations of the sdB primary star. The short-term orbit of $\tau_3$ has a relatively high determinacy 
because the observations have already covered about two cycles, while the long-term orbit of $\tau_{4}$ is currently 
preliminary because only about 67 \% of the 27-year period has been covered.

\section{Orbital Stability of Two-Planet System}

At this stage the most likely interpretation of the period change is that it is due to two planets, and this raises the question 
of their orbital time evolution. In recent times the orbital stability of proposed multi-body circumbinary systems detected from 
timing measurements has been studied by Horner et al. (2011), Hinse et al. (2012) and Hinse et al. (2014b, and references therein). 
An orbital stability check can be used as an independent test of the feasibility of our two-companion model. We have therefore 
carried out a stability study for the best-fit orbit parameters of NY Vir in Table 4, under the assumption that the combined mass 
(0.588 $M_{\odot}$) of the eclipsing pair can be considered as a central object and that the binary perturbations are negligible. 
Using the variable time-step RADAU algorithm implemented in the \texttt{MERCURY} integration package (Chambers 1999), we computed 
several batches of single-orbit integrations considering various initial conditions. The total integration time span was $10^6$ yr. 
In detail, we considered orbits within the errors obtained from the bootstrap-resampling technique. In addition giving 
the uncertainty in the time of pericenter passage we started the two planets from various initial mean longitudes. Furthermore, 
since their masses are minimum values we also considered various inclinations and scaled the masses of the companions accordingly. 

In most cases we found the system to exhibit large-scale orbital instabilities. Instability occurs either due to ejection 
from the system or any pair of body collide with each other. In the top panel of Figure 4, we show the time evolution of 
the semi-major axes and eccentricities of the two companions for the best-fit model (Table 4) demonstrating an ejection of 
the outer companion after $\simeq 850, 000$ yr. The result is in contrast with our expectation that the system should be stable 
following bounded orbits. However, while still considering best-fit parameters, we then explored the system's orbital time 
evolution for the upper limit of the outer companion's eccentricity and encountered stable/quasi-periodic orbits with no sign of 
chaotic dynamics. In an attempt to study the dynamics of the two-planet system in some more detail, we have numerically explored 
the phase space of the outer companion in a neighborhood of its best-fit orbit including the 1-$\sigma$ parameter errors obtained 
from the bootstrap-resampling technique. As an example we present the results of considering three (minimum, nominal best-fit 
and maximum) initial eccentricities within the orbital error range of the inner companion.

We calculated the life-time (Horner et al. 2011) and the MEGNO\footnote{Mean Exponential Growth Factor of Nearby Orbits} factor 
($\langle Y \rangle$) (Cincotta \& Sim{\'o} 2000; Go{\'z}dziewski, Bois \& Maciejewski 2001) over a large set of orbits in 
the semi-major axis$-$eccentricity space of the outer companion. For this task we used the \texttt{MECHANIC/MPI} orbit 
integration package (S{\l}onina et al. 2015). The results are shown in Figure 5. In each panel the parameter error box 
in the semi-major axis-eccentricity space of the outer companion is shown by a thick vertical and horizontal line centered 
on the nominal best-fit values. The left-side panels shows the degree of chaotic time evolution of the system for various 
initial conditions of the outer companion. Regions with colors $\langle Y\rangle \sim 2$ follow quasi-periodic orbits. All other 
orbits follow chaotic orbits. The right-side panels of Figure 5 shows the life-time of the system (determined in the event 
of ejection or collision). As shown earlier the best-fit orbit of the outer companion is embedded in a region of high chaoticity 
and relatively short system life-time.

As an example of a bounded orbit we have followed the dynamical time evolution of the initial condition shown by 
a black dot in Figure 5 located at a semi-major axis corresponding to the best-fit value of the outer companion's semi-major axis. 
The eccentricity of the outer companion was chosen to be 0.17 well within the uncertainty range of 0.20 (and 17\% larger than 
the best-fit value). The time evolution of the orbit is shown in the bottom panel of Figure 4 now demonstrating that the system 
is stable with bounded quasi-periodic orbits over $10^6$ yr with no sign of stochastic/random walk in the orbital elements 
during that time period. The corresponding LTT model is plotted as a dashed curve in the third panel of Figure 3. 
However, we would like to point out that the outer companion's orbit is not well characterised with the current dataset.

\section{SUMMARY AND DISCUSSION}

The successful fit to the eclipse times reveals that the orbital period of NY Vir has varied due to a combination of 
two periodic oscillations with cycle lengths of $P_3$=8.2 yr and $P_4$=27.0 yr and semi-amplitudes of $K_3$=6.9 s and 
$K_4$=27.3 d, respectively. The sinusoidal variations, shown in the second and third panels of Figure 3, can reasonably be 
interpreted as a pair of LTT effects driven by two substellar objects with minimum masses of $M_3 \sin i_3$=2.8 M$\rm_{Jup}$ 
and $M_4 \sin i_3$=4.5 M$\rm_{Jup}$. Assuming the circumbinary objects to be in the orbital plane ($i$=81$^\circ$) of 
the eclipsing pair, the masses of the third and fourth components would match to giant planets. The period ratio of 
$P_{3}$/$P_{4}$ implies that the two companions are located close to the 3:10 mean motion resonance. To the best of 
our knowledge, the result would be the fifth case when two circumbinary companions would be in or close to any kind of 
resonance (cf. Lee et al. 2012a). This suggests that a long-term gravitational interaction between the two planets would 
result in capture into the 3:10 resonant configuration.

Although all historical timings agree quite well with the calculated LTT effects as seen in Figure 3, we cannot 
{\it a priori} exclude the possibility that the period modulations in the eclipse timing diagram are due to 
magnetic activity cycles in the M-type secondary star. Therefore, we studied this alternate interpretation. 
With the modulation period ($P_{3,4}$) and amplitude ($K$) for the sinusoidal variations, model parameters were obtained 
from the formulae given by Applegate (1992). According to this theory, the secondary star with only 0.003 $L_\odot$ should 
exhibit the predicted luminosity variations of $\Delta L_{\rm rms,3}$ $\ga 0.0012 L_\odot$ and 
$\Delta L_{\rm rms,4}$ $\ga 0.0005 L_\odot$ for the short-term and long-term period variations, respectively, which is 
about 4.0 and 1.7 times larger than the $\Delta L_{\rm rms}/L_2 \sim 0.1$ level he proposed. Moreover, the timing analyses 
for HW Vir (Lee et al. 2009a) and HS 0705+6700 (Qian et al. 2009) indicated that it is difficult for the model to produce 
the perfectly smooth periodic variations of the sdB+M eclipsing pairs. These mean that Applegate's mechanism cannot 
contribute significantly to the observed period change of NY Vir. On the other hand, a single periodic variation 
could be attributed to apsidal motion in an elliptical orbit (cf. Parsons et al. 2014). However, previous studies for 
NY Vir suggest that the binary orbit is circular which in turn excludes timing variations from apsidal precession. 

Because all other mechanisms (Applegate effect, apsidal motion, and AML) of the period change can be ruled out, 
the most likely explanation of the variation is a pair of LTT effects driven by the presence of circumbinary companions. 
However, we found the two best-fit orbits to result in the ejection of the outer companion after some 800,000 yr. 
This time-span is considerably longer than the life-times of previously studied circumbinary multi-body systems 
(Horner et al. 2012a,b, 2013; Wittenmyer et al. 2012, 2013). With the limited span of timing measurements 
the outer companion orbit is still lacking a firm characterisation. We have therefore explored the neighborhood of 
the best-fit orbit and found several regions in phase space for which the system exhibits bounded orbits over $10^6$ yr. 
Future timing measurements will further characterise the outer companion and we conjecture that 
the best-fit eccentricity for the outer companion will increase to produce long-lived stable orbits. 
On the other hand, the root-mean-square (rms) scatter of timing measurements around our two-LTT model is 2.8 s. 
Compared with the semi-amplitude of the inner LTT orbit of 7 s, the detection of the inner companion is only marginal 
with a formal detection significance of 2.5$\sigma$. This fact should motivate obtaining precise timing measurements 
better than $\sim$3 s to identify and understand the proposed planetary system.

Up to now, the possible presences of circumbinary companions have been reported in five of 13 known HW Vir-type eclipsing 
systems including NY Vir: HW Vir (Lee et al. 2009a; Beuermann et al. 2012), HS 0705+6700 (Qian et al. 2009, 2013), 
HS 2231+2441 (Qian et al. 2010), and NSVS 14256825\footnote{A recent detailed study by Hinse et al. (2014b) indicates that 
the planets proposed in NSVS 14256825 are not feasible and the eclipse timing data are insufficient to reliably constrain 
the LTT model.} (Almeida et al. 2013). All the companions are likely either massive planets or brown dwarfs (BDs) based on 
their masses. How and when, then, did the circumbinary objects in the eclipsing system form? Typically, BDs are built up by 
fragmentation of a protostellar cloud, while planets originate from a protoplanetary disk left over from the protostellar disk 
of their newly formed host star (Udry \& Santos 2007). Perets (2010, 2011) suggested that a circumbinary planet orbiting 
the evolved binaries formed from protoplanetary disk material (first generation), from the secondary generation disk formed 
in the ejected envelope during post-MS evolution (second generation), or by the scattering of an S-type circumstellar planet 
into a circumbinary orbit. The second generation planets form in much older systems than the first generation ones and have 
a different source of material. If the circumbinary companions in the sdB+M eclipsing systems are formed in 
the first generation disk, their masses might be increased by accreting a large amount of the material replenished 
during the post-MS evolution. In consequence, more massive planets could become brown dwarf companions and be found in 
the BD dessert regime. At present, it is difficult to differentiate between the two types of substellar populations and 
additional information on such circumbinary objects is needed. The discovery and study of the circumbinary substellar objects 
around old evolved binary systems could shed new light on our understanding of both planet formation and binary evolution. 
In the case of the eclipsing system NY Vir, because the primary component is a pulsating sdB star like V391 Peg 
(Silvotti et al 2007), the pulsations can be used as an independent tool to confirm the circumbinary planets discovered 
by the eclipse timings.

\section*{Acknowledgments}
The authors wish to thank Dr. Maja Vu\v ckovi\'c for sending us their VLT/ULTRACAM photometric data. We also thank the staffs 
of LOAO for assistance with our observations. We appreciate the careful reading and valuable comments of the anonymous referee. 
This research has made use of the Simbad database maintained at CDS, Strasbourg, France, and was supported by the KASI 
(Korea Astronomy and Space Science Institute) grant 2014-1-400-06. Numerical simulations were carried out on the KMTNet 
computing cluster at KASI and the SFI/HEA Irish Centre for High-End Computing (ICHEC).

\clearpage
\begin{figure}
 \includegraphics[]{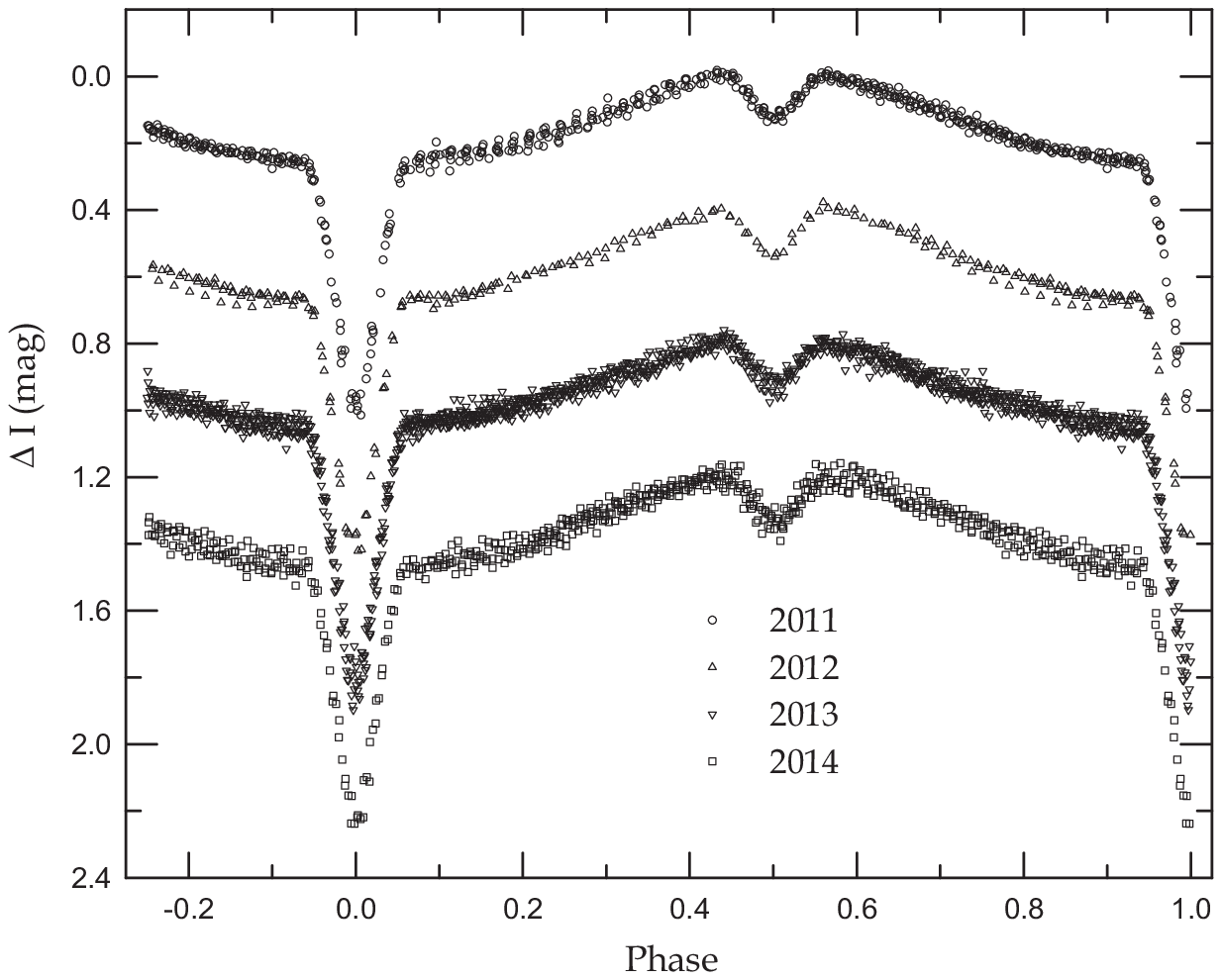}
 \caption{Light curves of NY Vir obtained at LOAO. The 2012--2014 curves are displaced vertically for clarity. }
 \label{Fig1}
\end{figure}

\begin{figure}
 \includegraphics[]{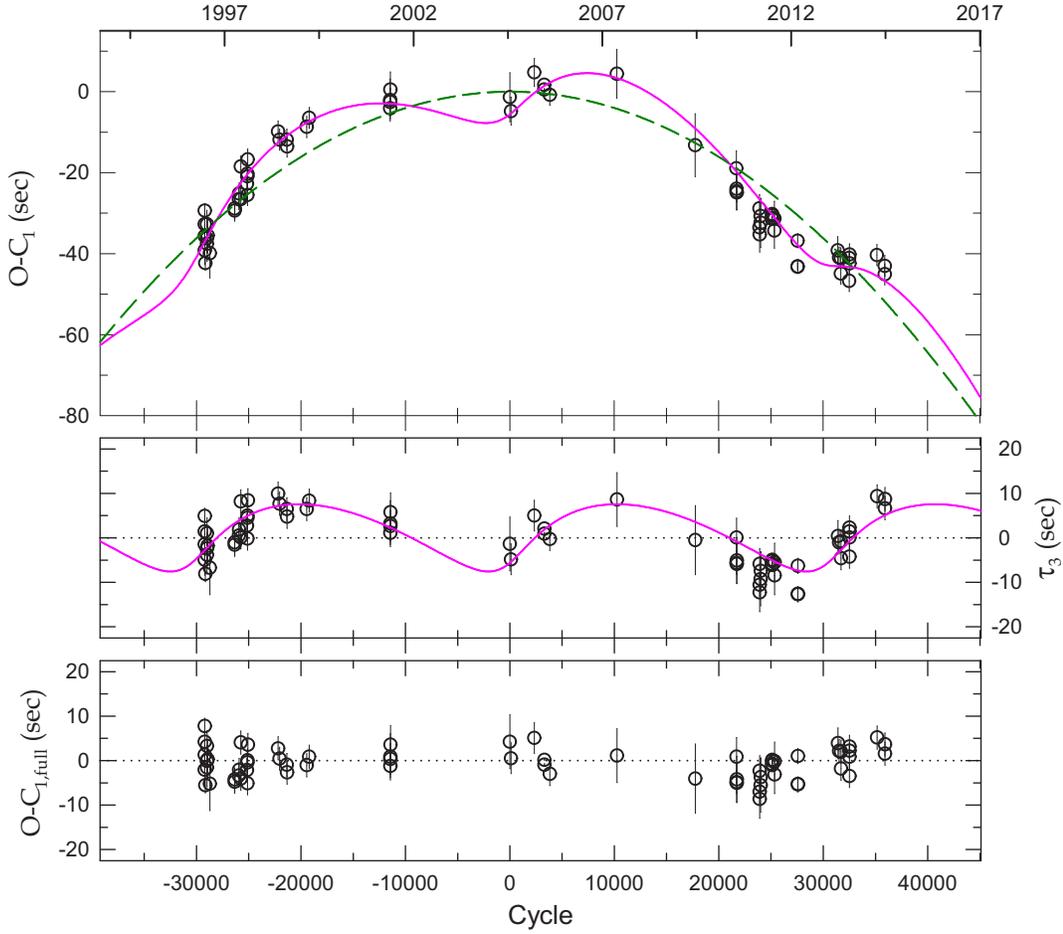}
 \caption{$O$--$C_1$ diagram of NY Vir constructed with the linear terms of the quadratic {\it plus} LTT ephemeris. In the top panel, 
 the solid curve and the dashed parabola represent the full contribution and the quadratic term of the equation, respectively. 
 The middle panel displays the LTT orbit with a cycle length of 8.42 yr and the bottom panel the residuals from the full ephemeris.}
 \label{Fig2}
\end{figure}

\begin{figure}
 \includegraphics[]{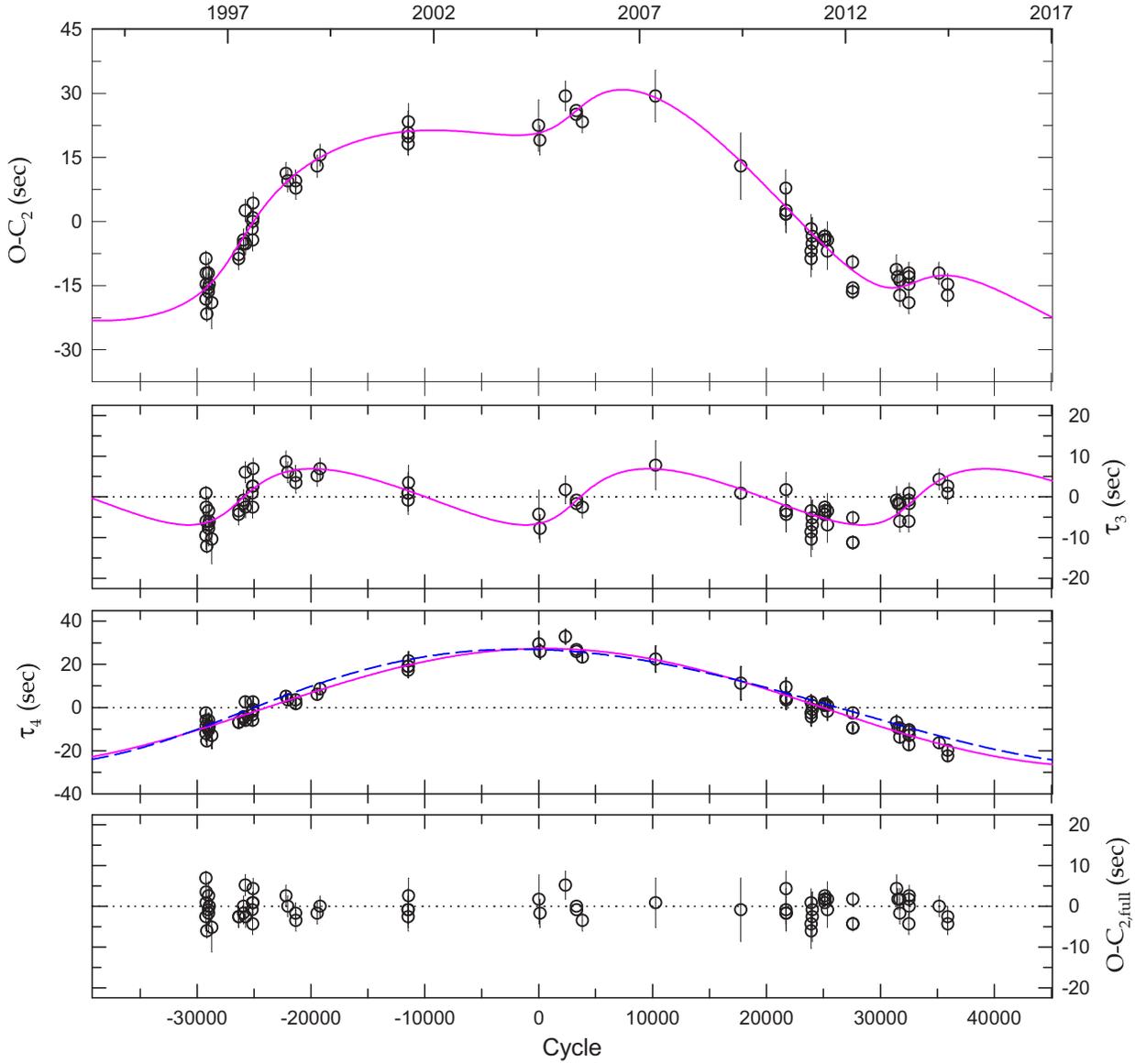}
 \caption{$O$--$C_2$ diagram of NY Vir with respect to the two-LTT ephemeris. In the uppermost panel, the continuous curve 
 represents the full effect of equation (2). The second and third panels display the short-term ($\tau_{3}$) and long-term 
 ($\tau_{4}$) LTT orbits, respectively. The bottom panel shows the residuals from the complete ephemeris. The dashed curve 
 in the third panel is the LTT model corresponding to the black dot ($e_4$=0.17) of Figure 5b. } 
\label{Fig3}
\end{figure}

\begin{figure}
\vbox{
\centerline{
 \includegraphics[scale=0.7]{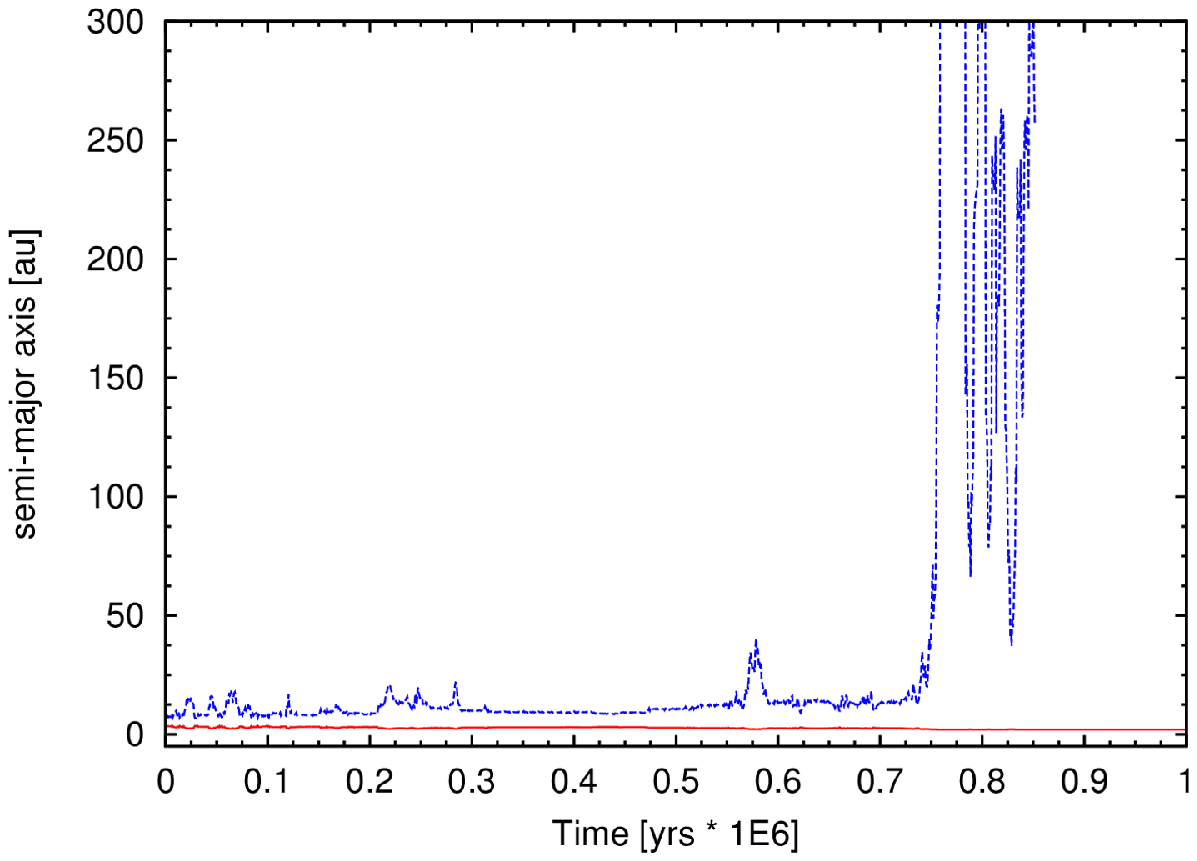}
 \includegraphics[scale=0.7]{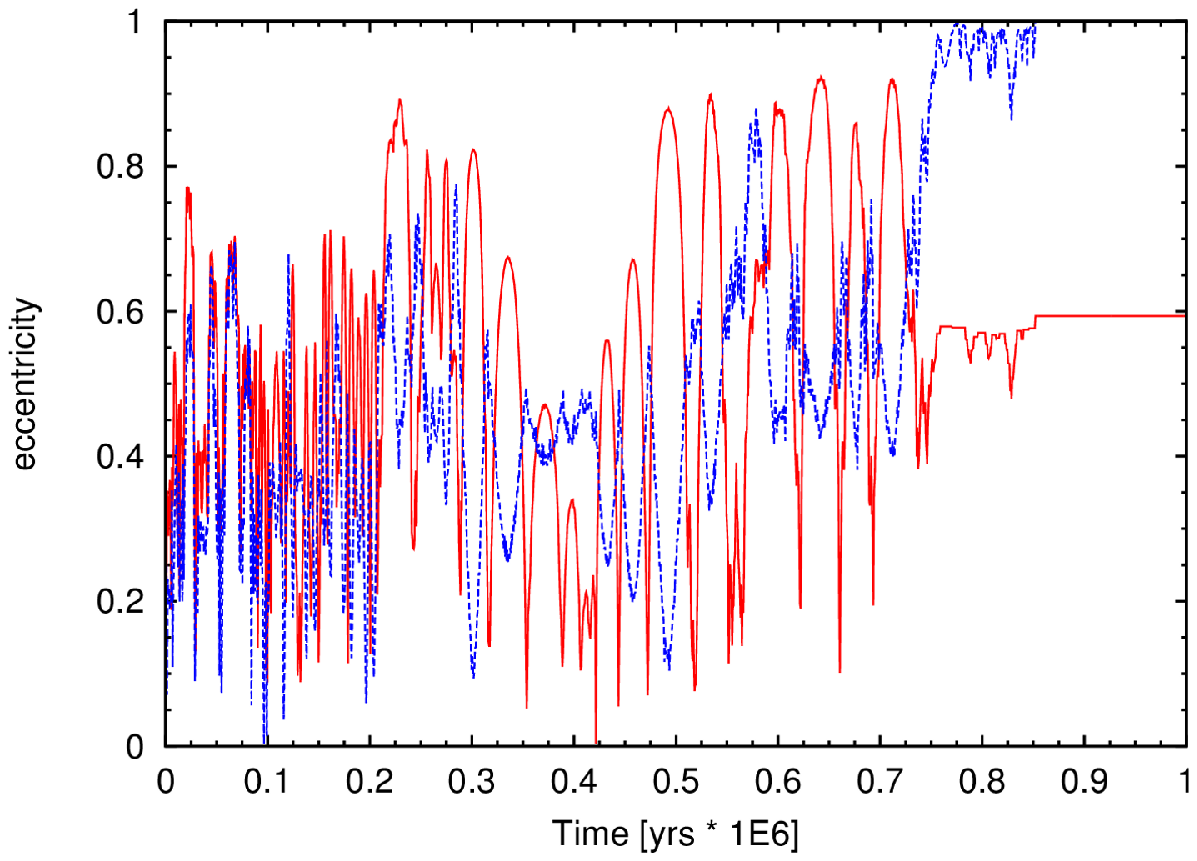}
}
}
\vskip 0.5cm 
\vbox{
\centerline{
 \includegraphics[scale=0.7]{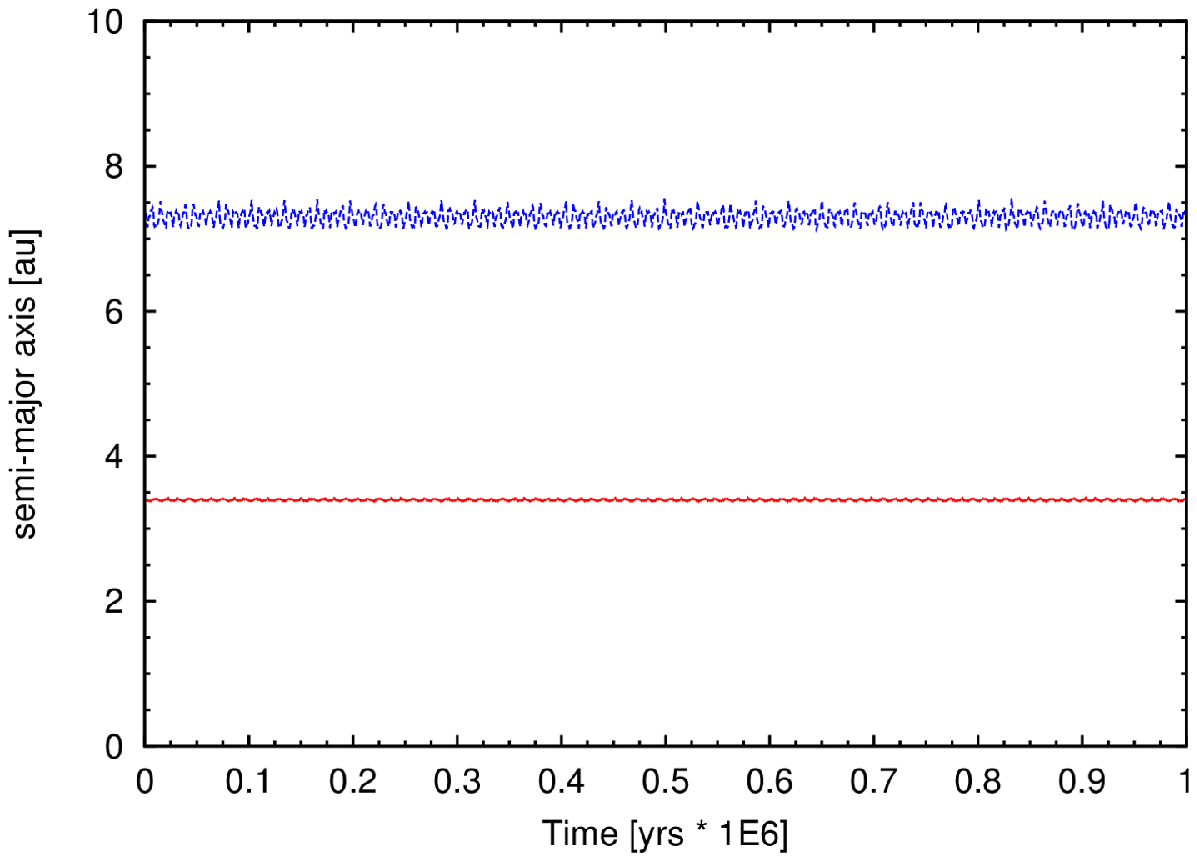}
 \includegraphics[scale=0.7]{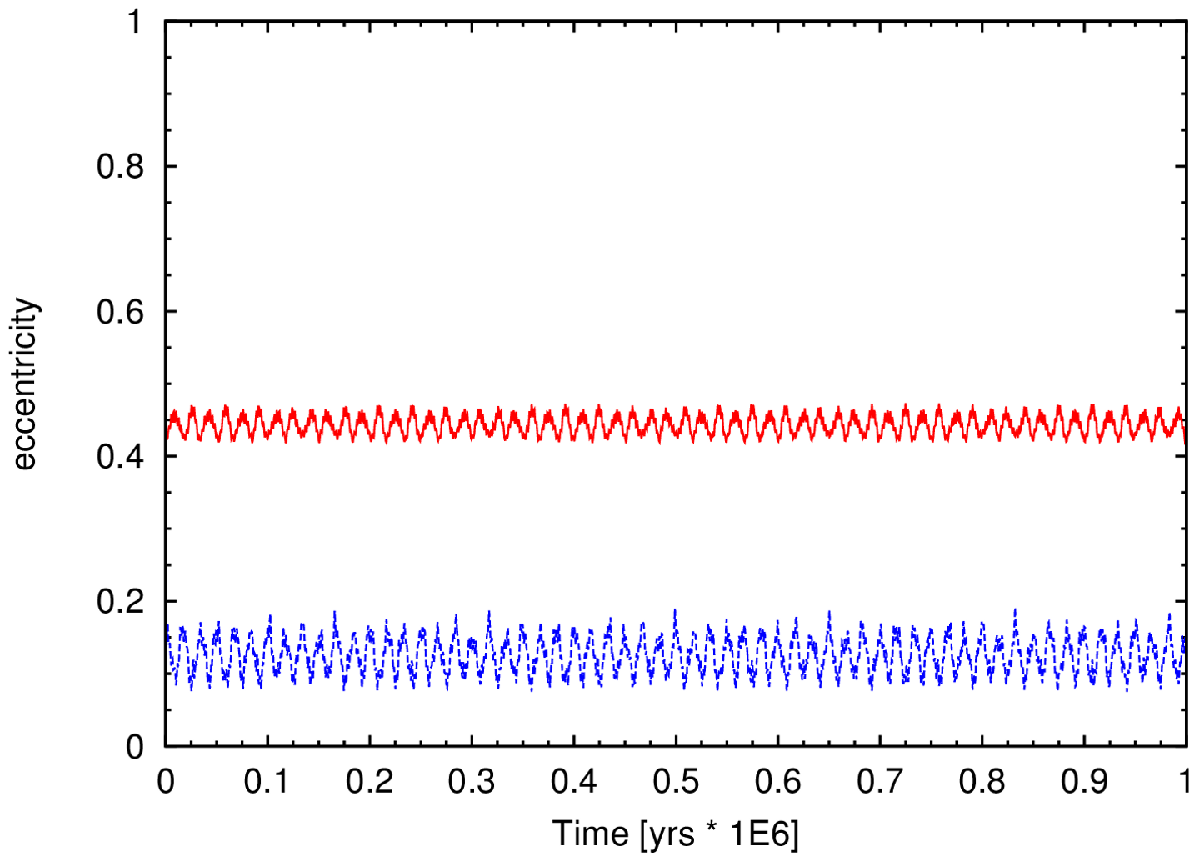}
}
}
 \caption{Orbital time evolution of the inner (red) and outer (blue) companions over $10^6$ yr using the RADAU algorithm. 
 \emph{Top panel:} Initial conditions are taken to be best-fit model parameters as shown in Table 4. The temporal evolution 
 clearly shows a strong gravitational interaction between the two proposed companions, resulting in the ejection of 
 the outer companion after 850,000 yr due to a mutual close encounter. \emph{Bottom panel:} Hypothetical case of  placing 
 the outer companion at the black dot shown in Figure 5b (middle panel). All other parameters are taken to be best-fit values 
 of the two-LTT model. The two companions are now perturbing each other on smaller magnitudes and the system follows 
 long-lived orbits over at least a one-million year time scale.}
\label{Fig4}
\end{figure}

\begin{figure*}
\vbox{
\centerline{
 \includegraphics[scale=0.7]{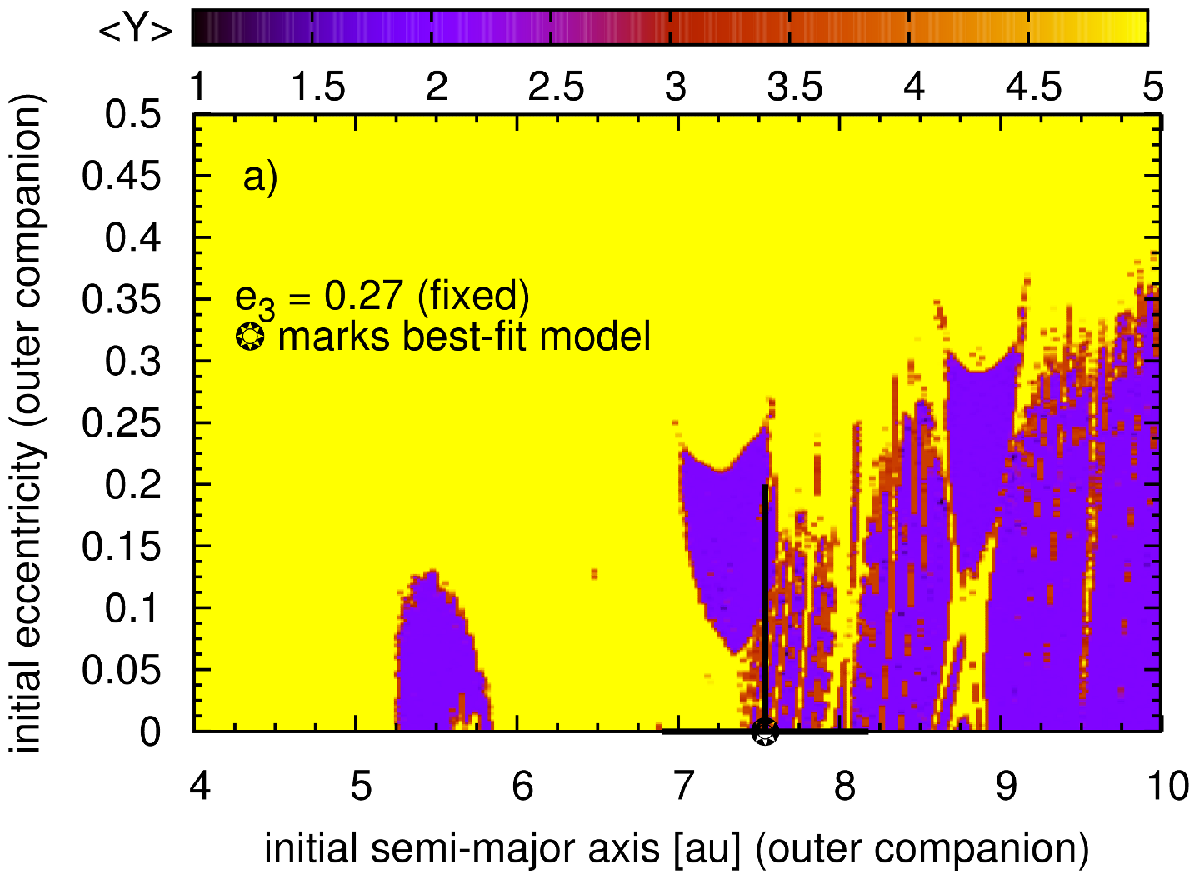}
 \includegraphics[scale=0.7]{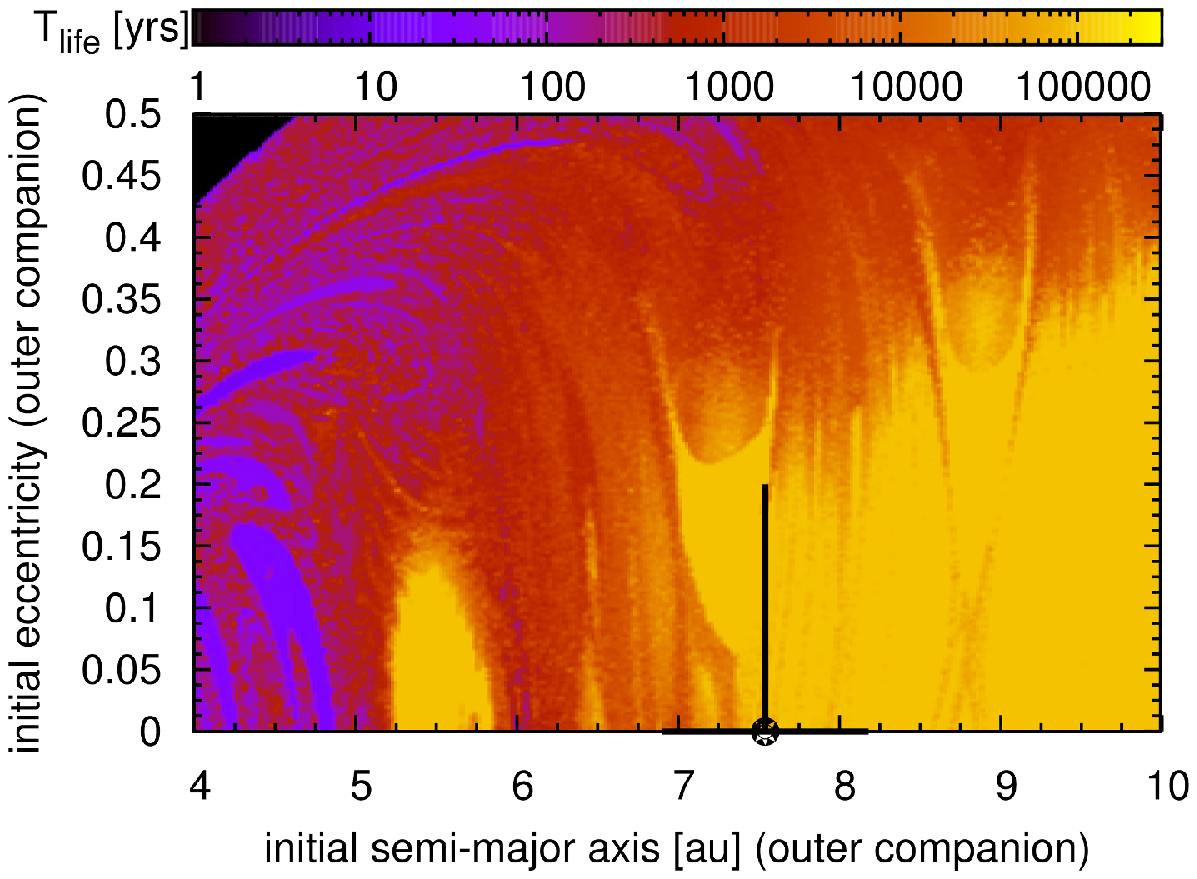}
}
}
\vskip 0.5cm 
\vbox{
\centerline{
 \includegraphics[scale=0.7]{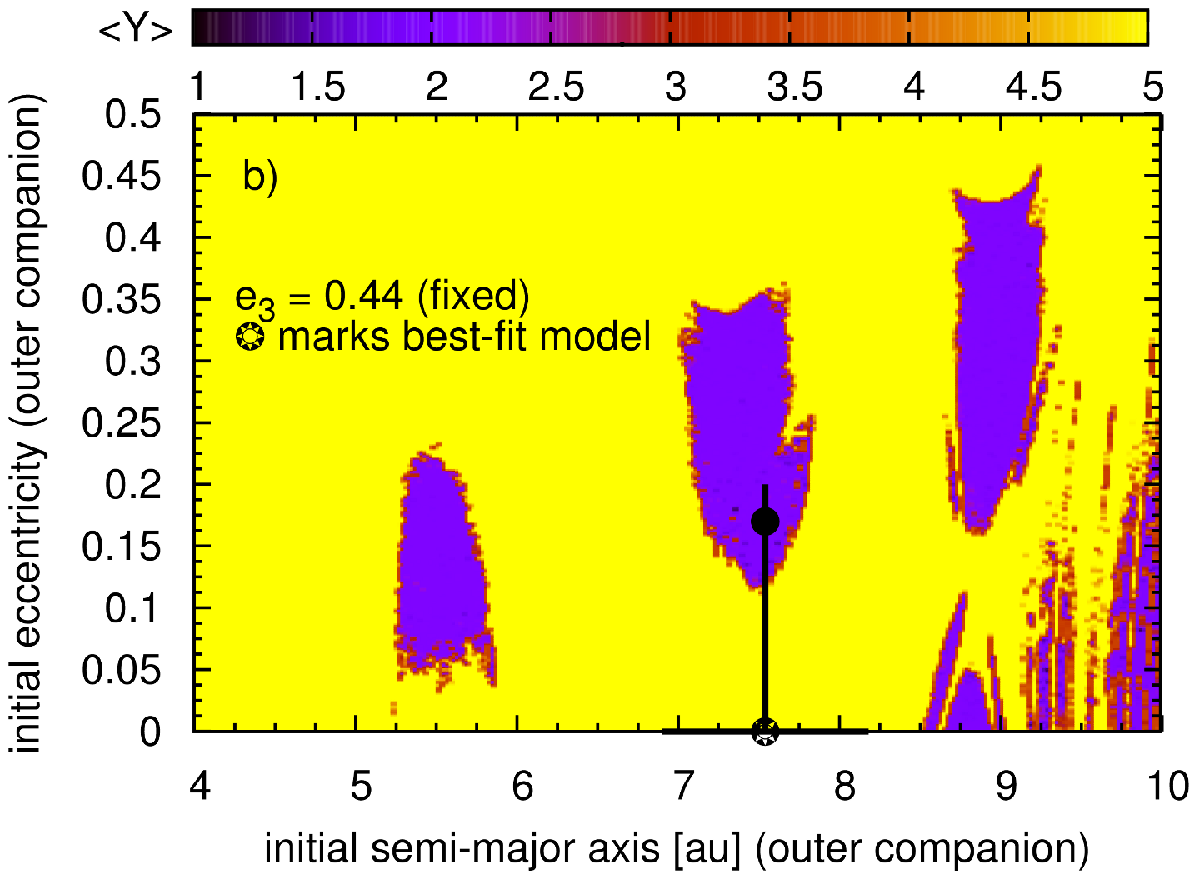}
 \includegraphics[scale=0.7]{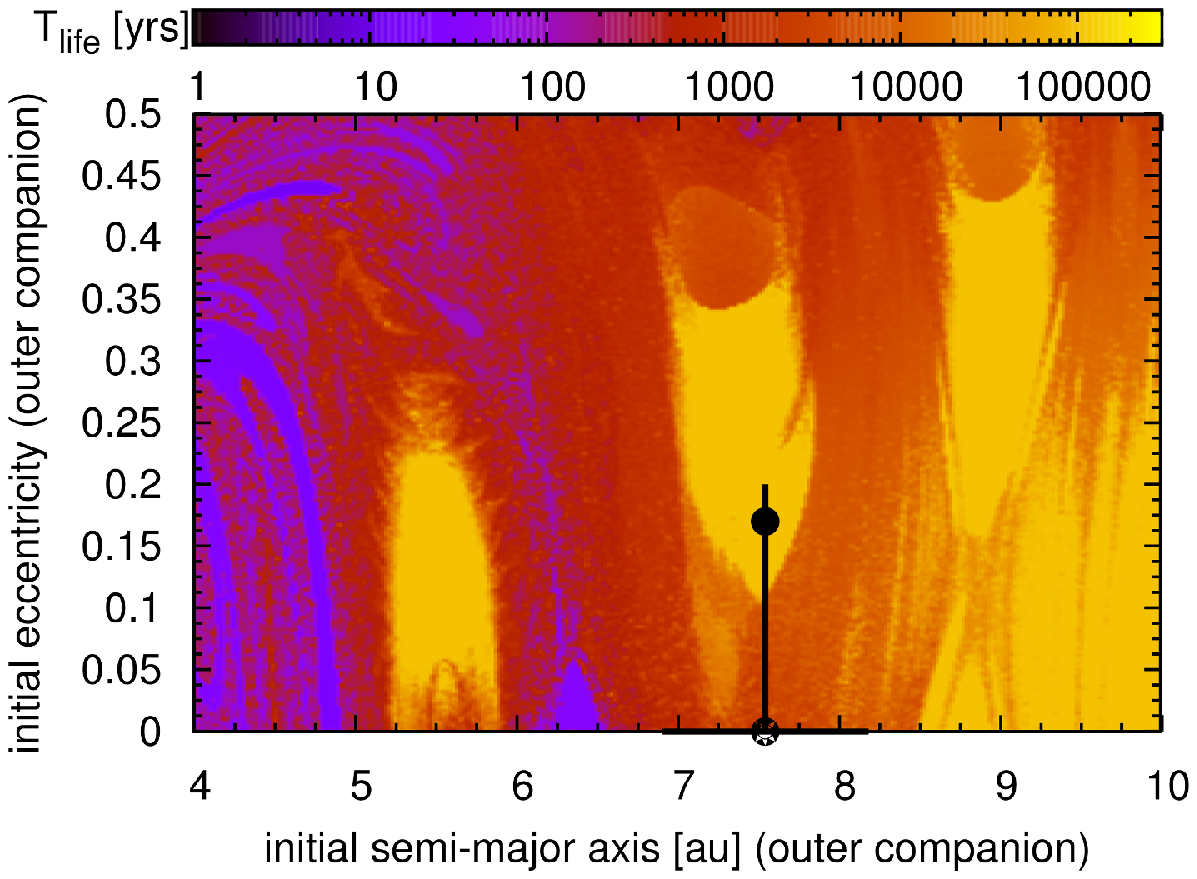}
}
}
\vskip 0.5cm 
\vbox{
\centerline{
 \includegraphics[scale=0.7]{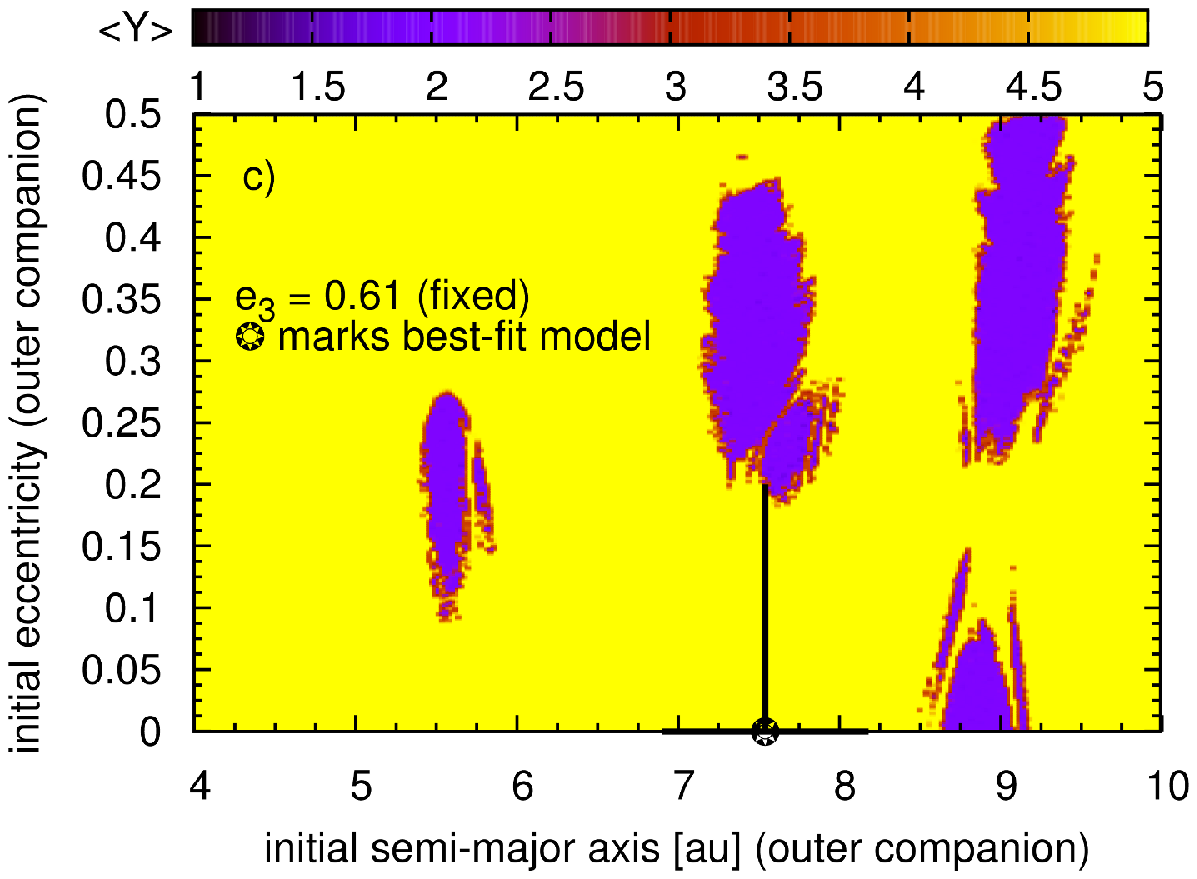}
 \includegraphics[scale=0.7]{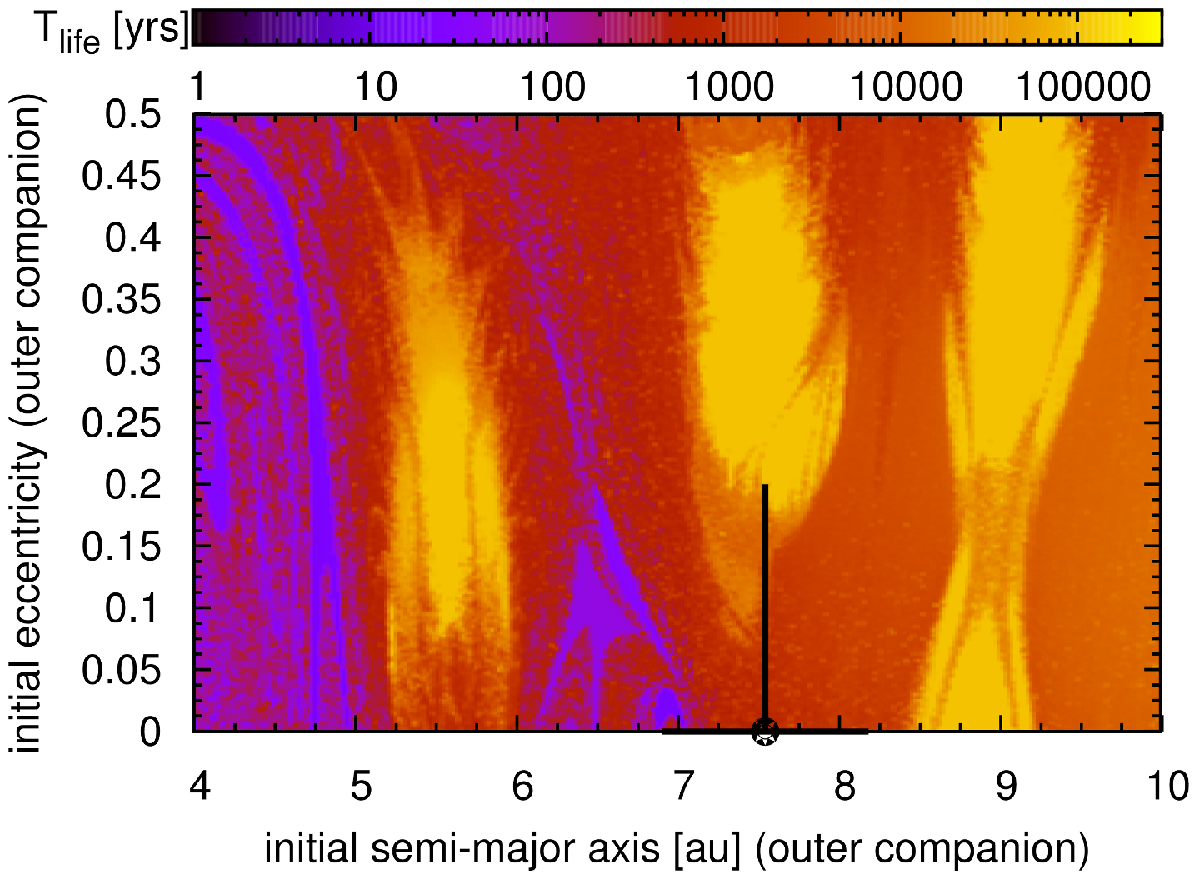}
}
}
 \caption{Dynamical maps and orbital parameter survey of the outer companion using the MECHANIC orbit integration package. 
 A given integration is terminated if the distance of any object reached 100 au or in the event of an ejection/collision or 
 for $\langle Y\rangle > 12$. The best-fit semi-major axis (7.54 au) is shown with a vertical line. The parameter uncertainty 
 in the semi-major axis and eccentricity of the outer companion is shown. The black dot indicates the location of 
 a hypothetical initial condition rendering the overall system to become stable and long-lived. \emph{Left-hand panels:} 
 MEGNO map of initial conditions in the neighborhood of the outer companion for three different eccentricities 
 (initially fixed to its minimum, nominal best-fit and maximum value) of the inner companion. Regions with quasi-periodic 
 (bounded) orbits result in $\langle Y \rangle \sim 2$ while strongly chaotic orbits have $\langle Y \rangle \gg 5$. 
 \emph{Right-hand panels:} Corresponding life-time map considering the same parameter range. A clear correlation between 
 quasi-periodic and long-lived orbits is evident.} 
\label{Fig5}
\end{figure*}

\clearpage
\begin{table}
\caption{Observing log of NY Vir. }
\begin{tabular}{lccccc}
\hline
Season   & Observing Interval  & $N_{\rm night}$  & CCD Type         & FOV (arcmin$^2$) & $N_{\rm obs}$   \\ 
\hline                                         
2011     & Jan. 28$-$Feb. 11   & 7                & FLI IMG4301E 2K  & 22$\times$22     &  468            \\  
2012     & Feb. 01$-$Feb. 02   & 2                & ARC 4K           & 28$\times$28     &  200            \\  
2013     & Feb. 23$-$Jun. 15   & 6                & ARC 4K           & 28$\times$28     &  1,133          \\  
2014     & Mar. 08$-$May  24   & 2                & ARC 4K           & 28$\times$28     &  569            \\  
\hline
\end{tabular}
\end{table}

\begin{table}
\caption{$I$-band photometry of NY Vir. A sample is shown here: the full version is provided as supplementary material to the online article. }
\begin{tabular}{lrc}
\hline
BJD               & Diff. Mag  & $\sigma_{\rm Mag}$  \\ 
\hline                                         
2,455,589.880291  & $-$0.0050    & 0.0070              \\
2,455,589.881251  &    0.0073    & 0.0072              \\
2,455,589.882212  &    0.0434    & 0.0073              \\
2,455,589.883289  &    0.0721    & 0.0064              \\
2,455,589.884481  &    0.1149    & 0.0064              \\
2,455,589.885673  &    0.1218    & 0.0066              \\
2,455,589.886865  &    0.1191    & 0.0064              \\
2,455,589.888069  &    0.0966    & 0.0064              \\
2,455,589.889261  &    0.0587    & 0.0060              \\
2,455,589.890454  &    0.0175    & 0.0064              \\
\hline
\end{tabular}
\end{table}

\begin{table}
\caption{New times of minimum light for NY Vir.}
\begin{tabular}{llcc}
\hline
BJD               & Error          & MIN  & Note                         \\ 
\hline                                
2,453,509.512963  & $\pm$0.000001  & I    & Vu\v ckovi\'c et al. (2007)  \\
2,453,509.563606  & $\pm$0.000001  & II   & Vu\v ckovi\'c et al. (2007)  \\
2,453,509.613966  & $\pm$0.000003  & I    & Vu\v ckovi\'c et al. (2007)  \\
2,453,509.664669  & $\pm$0.000003  & II   & Vu\v ckovi\'c et al. (2007)  \\
2,455,589.88599   & $\pm$0.00005   & II   & LOAO                         \\
2,455,589.93642   & $\pm$0.00003   & I    & LOAO                         \\
2,455,589.98693   & $\pm$0.00004   & II   & LOAO                         \\
2,455,590.03749   & $\pm$0.00004   & I    & LOAO                         \\
2,455,590.94656   & $\pm$0.00005   & I    & LOAO                         \\
2,455,590.99701   & $\pm$0.00017   & II   & LOAO                         \\
2,455,599.88685   & $\pm$0.00011   & II   & LOAO                         \\
2,455,599.98762   & $\pm$0.00008   & II   & LOAO                         \\
2,455,600.03803   & $\pm$0.00007   & I    & LOAO                         \\
2,455,600.89677   & $\pm$0.00014   & II   & LOAO                         \\
2,455,601.04821   & $\pm$0.00005   & I    & LOAO                         \\
2,455,604.02814   & $\pm$0.00007   & II   & LOAO                         \\
2,455,958.94764   & $\pm$0.00002   & I    & LOAO                         \\
2,455,958.99821   & $\pm$0.00008   & II   & LOAO                         \\
2,455,959.04873   & $\pm$0.00002   & I    & LOAO                         \\
2,455,959.95780   & $\pm$0.00001   & I    & LOAO                         \\
2,455,960.00786   & $\pm$0.00013   & II   & LOAO                         \\
2,456,346.89952   & $\pm$0.00006   & II   & LOAO                         \\
2,456,346.95002   & $\pm$0.00004   & I    & LOAO                         \\
2,456,358.92060   & $\pm$0.00018   & II   & LOAO                         \\
2,456,358.97090   & $\pm$0.00003   & I    & LOAO                         \\
2,456,359.02188   & $\pm$0.00008   & II   & LOAO                         \\
2,456,376.85068   & $\pm$0.00003   & I    & LOAO                         \\
2,456,376.90136   & $\pm$0.00019   & II   & LOAO                         \\
2,456,376.95174   & $\pm$0.00004   & I    & LOAO                         \\
2,456,377.00181   & $\pm$0.00008   & II   & LOAO                         \\
2,456,457.66350   & $\pm$0.00002   & I    & LOAO                         \\
2,456,457.71401   & $\pm$0.00010   & II   & LOAO                         \\
2,456,457.76445   & $\pm$0.00003   & I    & LOAO                         \\
2,456,458.67367   & $\pm$0.00003   & I    & LOAO                         \\
2,456,458.72452   & $\pm$0.00008   & II   & LOAO                         \\
2,456,458.77466   & $\pm$0.00003   & I    & LOAO                         \\
2,456,724.90146   & $\pm$0.00015   & II   & LOAO                         \\
2,456,724.95176   & $\pm$0.00003   & I    & LOAO                         \\
2,456,725.00204   & $\pm$0.00016   & II   & LOAO                         \\
2,456,801.67332   & $\pm$0.00009   & II   & LOAO                         \\
2,456,801.72384   & $\pm$0.00003   & I    & LOAO                         \\
2,456,801.77407   & $\pm$0.00012   & II   & LOAO                         \\
2,456,801.82488   & $\pm$0.00003   & I    & LOAO                         \\
\hline
\end{tabular}
\end{table}

\begin{table}
\caption{Parameters for the LTT orbits of NY Vir.}
\begin{tabular}{lccccc}
\hline
Parameter                & Quadratic {\it plus} LTT               && \multicolumn{2}{c}{Two-LTT}                            & Unit            \\ [1.5mm] \cline{4-5} \\ [-2.0ex]
                         & $\tau_{3}$                             && $\tau_{3}$                & $\tau_{4}$                 &                 \\ 
\hline                                                             
$T_0$                    & 2,453,174.4429776(62)                  && \multicolumn{2}{c}{2,453,174.442699(91)}               & BJD             \\
$P$                      & 0.10101596816(37)                      && \multicolumn{2}{c}{0.1010159668(43)}                   & d               \\
$a_{12}\sin i_{3,4}$     & 0.0160(13)                             && 0.0153(22)                & 0.055(16)                  & au              \\
$e$                      & 0.46(18)                               && 0.44(17)                  & 0.00(20)                   &                 \\
$\omega$                 & 314.0(3.0)                             && 346(15)                   & 333(15)                    & deg             \\
$n$                      & 0.1171(33)                             && 0.1205(27)                & 0.0365(50)                 & deg d$^{-1}$    \\
$T$                      & 2,453,167(40)                          && 2,453,472(141)            & 2,450,031(497)             & BJD             \\
$P_{3,4}$                & 8.42(23)                               && 8.18(18)                  & 27.0(3.7)                  & yr              \\
$K$                      & 7.57(60)                               && 6.9(1.0)                  & 27.3(8.2)                  & s               \\
$f(M_{3,4})$             & 5.79(49)$\times 10^{-8}$               && 5.34(76)$\times 10^{-8}$  & 2.25(74)$\times 10^{-7}$   & M$_\odot$       \\
$M_{3,4} \sin i_{3,4}$   & 2.85(11)                               && 2.78(19)                  & 4.49(72)                   & M$\rm_{Jup}$    \\
$a_{3,4} \sin i_{3,4}$   & 3.457(69)                              && 3.39(12)                  & 7.54(64)                   & au              \\[0.5mm]
$A$                      & $-$4.65(12)$\times 10^{-13}$           &&                           &                            & d               \\
$dP$/$dt$                & $-$3.36(10)$\times 10^{-9}$            &&                           &                            & d yr$^{-1}$     \\[0.5mm]
rms scatter              & 3.39                                   && \multicolumn{2}{c}{2.83}                               & s               \\
$\chi^2_{\rm red}$       & 1.31                                   && \multicolumn{2}{c}{1.09}                               &                 \\
\hline
\end{tabular}
\end{table}

\bsp
\label{lastpage}
\end{document}